\documentclass{aa}  
\usepackage{graphicx}
\pdfoutput=1
\usepackage{txfonts}
\usepackage{natbib} 
\usepackage{float}
\usepackage{rotating}
\usepackage{amsmath}
\bibpunct{(}{)}{;}{a}{}{,} 
\begin{document}
   \title{Disk evolution in the solar neighborhood}
   \subtitle{I. Disk frequencies from 1 to 100 Myr}

   \author{\'Alvaro Ribas\inst{\ref{esac},\ref{cab},\ref{insa}}
     \and
     Bruno ~Mer\'{i}n\inst{\ref{herschel}}
     \and
     Herv\'e~Bouy\inst{\ref{cab}} 
     \and
     Luke T. Maud \inst{\ref{leeds}}
     }

   \offprints{\'A. Ribas}
	\institute{
	European Space Astronomy Centre (ESA), P.O. Box, 78, 28691 Villanueva de la Ca\~{n}ada, Madrid, Spain\label{esac} \\
              \email{aribas@cab.inta-csic.es}
         \and
        Centro de Astrobiolog\'{i}a, INTA-CSIC, P.O. Box - Apdo. de correos 78, Villanueva de la Ca\~nada Madrid 28691, Spain\label{cab} 
        \and
	Ingenier\'ia y Servicios Aeroespaciales-ESAC, P.O. Box, 78, 28691 Villanueva de la Ca\~{n}ada, Madrid, Spain\label{insa}
        \and
	Herschel Science Centre, ESAC-ESA, P.O. Box, 78, 28691 Villanueva de la Ca\~{n}ada, Madrid, Spain\label{herschel}
        \and
        School of Physics \& Astronomy, EC Stoner Building, University of Leeds, Leeds, LS2 9JT, West Yorkshire, England \label{leeds}}
   \date{Received 3 September 2013; Accepted 13 November 2013}

 
  \abstract
{}
   {We study the evolution of circumstellar disks in 22 young (1 to 100~Myr) nearby (within 500~pc) associations over the entire mass spectrum using photometry covering from the optical to the mid-infrared.}
   {We compiled a catalog of 2\,340 spectroscopically-confirmed members of these nearby associations. We analyzed their spectral energy distributions and searched for excess related to the presence of protoplanetary disks. The dataset has been analyzed in a homogeneous and consistent way, allowing for meaningful inter-comparison of results obtained for individual regions. Special attention was given to the sensitivity limits and spatial completeness of the observations.}
   {We derive disk fractions as probed by mid-infrared excess in the 22 regions. The unprecedented size of our sample allows us to confirm the timescale of disk decay reported in the literature and to find new trends. The fraction of excess sources increases systematically if measured at longer wavelengths. Disk percentages derived using different wavelength ranges should therefore be compared with caution. The dust probed at 22--24\,$\mu$m evolves slower than that probed at shorter wavelengths (3.4--12~$\mu$m). Assuming an exponential decay, we derive a timescale  $\tau=$4.2$\sim$5.8~Myr at 22--24~$\mu$m for primordial disks, compared to 2$\sim$3~Myr at shorter wavelength (3.4--12~$\mu$m). 
Primordial disks disappear around 10$\sim$20~Myr. Their decline matches in time a brief increase of the number of ``evolved'' disks (defined here as including transitional and debris disks).  There is more dispersion in the fraction of excess sources with age when measured at 22--24~$\mu$m in comparison to shorter wavelengths.}
   {The increase in timescale of excess decay at longer wavelength is compatible with inside-out disk clearing scenarios. The increased timescale of decay and larger dispersion in the distribution of disk fractions at 22--24~$\mu$m suggest that the inner (terrestrial-planet forming) and outer (giant-planet forming) zones evolve differently, the latter potentially following a variety of evolutionary paths. The drop of primordial disks and the coincident rise of evolved disks at 10~Myr are compatible with planet formation theories suggesting that the disappearance of the gas is immediately followed by the dynamical stirring of the disk.}


   \maketitle


\section{Introduction}

The structure and evolution of protoplanetary disks provide crucial information to better understand stellar and planetary formation \citep[see][for a complete review on different aspects of protoplanetary disks evolution]{Williams2011}. The timescale of disk evolution, and its dependence on stellar age, mass, environment and multiplicity, is a fundamental parameter to constrain stellar formation theories and numerical simulations. The inter-phase between protoplanetary and debris disks is also key to understand the first stages of planetary formation. However, details of the processes involved in the dispersal of protoplanetary disks and the transition to debris disks have been elusive mostly because of the long distances to such disks (hence difficulty to resolved them) and the size of the samples studied so far. 

Circumstellar disks are often detected via infrared (IR) excess over the stellar photosphere: dust in the disk is heated by stellar radiation, and re-emits it in the mid/far-IR regime. The advent of IR observatories such as AKARI, the \textit{Spitzer } Space Observatory, and more recently the {\it Herschel} Space Observatory produced very sensitive surveys over vast areas in several young nearby associations and clusters \citep{Evans2009,Ishihara2010}. Combined with all-sky or wide field surveys in the optical (e.g CMC~14, and SDSS) and in the near/mid-IR (e.g DENIS, UKIDSS, and WISE), they provide a sensitive and homogeneous photometric coverage of a large number of nearby regions and make a new and unique database for the study of protoplanetary disks.

We present a sample of 2\,340 young stars and brown dwarfs members of 22 nearby ($<$500~pc) associations (13 star-forming regions and 9 young nearby associations) with ages ranging from 1\, to\,100 Myr. This sample of unprecedented size is a compilation of 37 spectroscopic studies published in the literature over the past two decades. The spectral energy distribution (SED) of each source was obtained by merging various catalogs covering the wavelength range from the optical ($\sim$V-band) to the mid-IR (22$\sim$24~$\mu$m). The quantity and the quality of the photometric measurements allow us to derive and study disk fractions in a consistent way for all the regions. 

The present manuscript is the first of a series of articles where disk evolution will be investigated with a large and representative sample of young stars in the solar neighborhood. In Sect.~2, we present the sample and its characteristics. We describe the astrometric and photometric quality assessment and data curation process in Sect.~3. The SED fitting procedure is described in Sect.~4. In Sect.~5, we describe the method and criterion adopted to identify mid-IR excesses and the method used to derive excess (disk) fractions. The results are given in Sect.~6, and discussed in Sect.~7.

\section{Sample}\label{sample}
We compiled a sample of spectroscopically confirmed members of nearby regions using catalogs from the literature for the 25~Orionis, Chameleon~I \& II, Corona Australis, IC~348, $\lambda$-Orionis, Lupus, NGC~1333, Ophiuchus, $\sigma$-Orionis, Serpens, Taurus, and Upper Scorpius associations and clusters. This original sample was then completed by adding the  spectroscopically confirmed members of young nearby associations from the SACY survey presented in \citet{Torres2008}, which includes the AB~Doradus, Argus, $\beta$-Pictoris,  Carina, Columba, $\eta$-Chamaeleon, Octantis, Tuc-Hor (THA), and TW~Hya (TWA) associations. Table~\ref{tab:sample} summarizes of the basic properties of these regions, as well as the references used for the spectroscopically confirmed samples. 

Young nearby associations offer several advantages for the study of stellar and planetary formation. Their distances and relative ages are in general relatively well known. Additionally, their proximity and youth make the detection of their least massive members easier, ensuring a complete (or at least representative) census of their stellar content from the most massive members to the substellar and planetary mass objects. As a result, they have extensively been studied over the past four decades \citep[see e.g][ and references therein]{Reipurth2008N, Reipurth2008S}. Additionally, nearby associations with ages in the range 5\,-\,80\,Myr provide a good opportunity for studying the transition from protoplanetary to debris disks.

From the various catalogs found in the literature (see Table~\ref{tab:sample}), we chose to keep only those objects with spectroscopically-obtained spectral type . This choice was motivated by two main reasons:
\begin{itemize}

\item Spectroscopic confirmation minimizes contamination by unrelated foreground or background sources. Photometric surveys are known to include as much as 30\% of contaminants \citep[see e.g ][]{Oliveira2009}.

\item Knowing the spectral type breaks the degeneracy between T$_{\rm eff}$ and visual extinction ($A_V$) in the SED fitting, making the $A_{\rm V}$ estimation more robust.

\end{itemize}

A total of 2\,627 sources fulfill this first requirement.

\begin{table*}
  \caption{Young nearby regions (top) and associations (bottom) included in this study. }
  \label{tab:sample}
  \begin{center}
    \begin{tabular}{lccccc}
      \hline\hline Name & Age & Distance  & Membership and SpT & \textit{Spitzer} photometry & Number of sources\\
      & (Myr) & (pc)        &              &   &       \\
      \hline
      25~Orionis & 7-10 & 330 & (1) & (1) & 46\\
      Cha~I  & 2 & 160-165 & (2) (3) (4) & (3) (4) & 212\\
      Cha~II & 2$\pm$2& 178$\pm$18 & (5) & (38) & 47\\
      CrA     & 1-3 & 138$\pm$16 & (6) (7) (8) (9) & (39) & 35\\
      IC~348  & 2 & 285-315 & (10) (11) & (38) (40) & 298\\
      $\lambda$ Orionis  &  4 & 400$\pm$40 & (12) & (41) & 114\\
      Lupus  & 1-1.5 &140 - 200 & (13) (14) (15) (16) & (38) & 217\\
      NGC~1333  & 1 & 235$\pm$18 & (17) & (42) & 74\\
      Ophiuchus  & 2-5& 120-145 & (18) (19) (20) (21) & (38) & 258\\
      $\sigma$ Orionis  & 2-3 & 440$\pm$30 & (22) (23) (24) (25) (26) (27) (28) & (43) & 104\\
      Serpens & 2 & 230$\pm$20 &  (17) (29) & (38) & 142\\
      Taurus & 1-2 & 140 &  (30) (31) (32) (33) & (32) (33) & 265\\
      Upper Sco & 11$\pm$2 & 140 & (34) (35) (36) & (35) & 405
      \\ \hline
      AB~Dor        &   50-100    & 34            & (37) &  \ldots & 92\\
      Argus       &   40                 & 106           & (37) &  \ldots & 51\\
      $\beta$ Pic   &   10                 & 31            & (37) &  \ldots & 54\\
      Carina      &   30                 & 85            & (37) &  \ldots & 37\\
      Columba       &   30                 & 82            & (37) &  \ldots & 58\\
      $\eta$ Cha    &   4-9                  & 108           & (37) &  \ldots & 27\\ 
      Octantis    &   20                 & 141           & (37) &  \ldots & 17\\
      Tuc-Hor    &   30                 & 48            & (37) &  \ldots & 49\\
      TW~Hya       &   8                  & 48            & (37) &  \ldots & 25\\
      \hline
    \end{tabular}

\tablefoot{Ages and distances are from \citet{Reipurth2008N} and \citet{Reipurth2008S} unless specified. We also tabulate references to the different studies used to compile the sample of objects and to the \textit{Spitzer} photometry, as well as the number of sources for each region.}

    \tablebib{
      (1) \citet{Hernandez2007_25Ori}; 
      (2) \citet{Luhman2007}; 
      (3) \citet{Luhman2008a}; 
      (4) \citet{Luhman2008b}; 
      (5) \citet{Spezzi2008}; 
      (6) \citet{Neuhauser2000}; 
      (7) \citet{Nisini2005}; 
      (8) \citet{SiciliaAguilar2008}; 
      (9) \citet{SiciliaAguilar2011}; 
      (10) \citet{Luhman2003}; 
      (11) \citet{AlvesdeOliveira2013}; 
      (12) \citet{Bayo2011}; 
      (13) \citet{Krautter1997}; 
      (14) \citet{Allen2007}; 
      (15) \citet{Comeron2008}; 
      (16) \citet{Mortier2011}; 
      (17) \citet{Winston2009}; 
      (18) \citet{Natta2002}; 
      (19) \citet{Wilking2005}; 
      (20) \citet{AlvesdeOliveira2010}; 
      (21) \citet{Erickson2011}; 
      (22) \citet{ZapateroOsorio2002}; 
      (23) \citet{Muzerolle2003}; 
      (24) \citet{Barrado2003}; 
      (25) \citet{Franciosini2006}; 
      (26) \citet{Caballero2007AA466}; 
      (27) \citet{Sacco2008}; 
      (28) \citet{Rigliaco2012}; 
      (29) \citet{Oliveira2009}; 
      (30) \citet{Luhman2004Taurus}; 
      (31) \citet{Monin2010}; 
      (32) \citet{Rebull2010}; 
      (33) \citet{Rebull2011}; 
      (34) \citet{Preibisch2002}; 
      (35) \citet{Carpenter2006}; 
      (36) \citet{Lodieu2011}; 
      (37) \citet{Torres2008}; 
      (38) \citet{Evans2009};
      (39) \citet{Peterson2011}; 
      (40) \citet{Lada2006}; 
      (41) \citet{Barrado2007}; 
      (42) \citet{Gutermuth2008}; 
      (43) \citet{Luhman2008SOri} 
    }

  \end{center}
\end{table*}

\subsection{Ages and distances}

Distances can be measured with different techniques, leading to several (sometimes inconsistent) values in the literature for the same region. With the exception of the SACY sample, most regions considered here are further away than 100\,pc, and hence beyond \textsc{Hipparcos}' sensitivity limit \citep{Perryman1997}. Ages are difficult to determine, and star formation is probably not an abrupt event but extended in time \citep[see e.g.][]{Hartmann2001}. Absolute ages are usually uncertain and sometimes controversial \citep{Bell2013}, but the \emph{relative} age sequence is expected to be accurate enough for the aim of our study and to perform a meaningful comparison of disk properties at different evolutionary stages (Soderblom et al. PPVI review, in prep). 

For consistency, we chose to use ages and distances provided in \citet{Reipurth2008N} and \citet{Reipurth2008S} except when updated measurements were found in the literature:
\begin{itemize}
\item Age of Cha~II from \citet{Spezzi2008},
\item Distance to NGC~1333 from \citet{Hirota2008},
\item Age and distance to Upper Scorpius from \citet{Pecaut2012},
\item Age and distance to 25~Orionis come from \citet{Hernandez2007_25Ori}. 
\end{itemize}

In the rest of our analysis, a conservative 50\,\% uncertainty level was assumed for the ages, unless a different estimate was provided in the literature. 

\subsection{Mass range covered by the sample}\label{sect:mass_coverage}

The completeness level of the sample (in terms of mass or luminosity) is critical to assess the relevance of any statistical result, as an incomplete sample would yield biased results. This issue becomes even more important when mixing catalogs from different sources, which may have different completeness levels.

All the star-forming regions selected in this study have been extensively studied in the past \citep[see i.e.][]{Reipurth2008N,Reipurth2008S}. The catalogs used to compile the sample presented here come from stellar population-oriented studies whose main objective was to define representative samples of members and derive accurate mass functions. On the high mass end, the various studies used in this analysis are expected to be complete since massive stars are bright and have been known for long. On the low mass end, the stellar population studies used in this analysis are in general complete or representative down the mid or late-M dwarfs. We therefore expect that our study is representative over the luminosity range between the most massive members (typically O, B or A stars depending on the association) and the mid/late-M dwarfs.

\section{Astrometric and photometric quality assurance}

\subsection{Astrometry}
In the past, astrometric calibration was more complicated because of the lack of accurate reference astrometric catalogs. As a result, some of the oldest surveys used in this study have a poor astrometric accuracy. We undertook a data healing procedure to ensure consistent astrometric accuracy for all objects across the different regions and associations. 

For the entire sample, we chose to use the 2MASS astrometry as reference \citep{2MASS}. 2MASS offers the advantage of homogeneously coverage of the entire sky, with an average astrometric accuracy of $\sim$0\farcs1. In this process, each target is associated to the closest 2MASS counterpart within 3\arcsec. We a posteriori verify that the mean separation between the original and the 2MASS coordinates is smaller than 0\farcs4, except in the case of Ophiuchus \citep[from][]{Erickson2011} and Upper Sco \citep[from][]{Preibisch2002} where systematic offsets larger than 1\arcsec\, with respect to the 2MASS catalog were found and corrected.  

Only 89 of the 2\,627 objects had no counterpart in 2MASS within 3\arcsec\, due to luminosities beyond the sensitivity limit of the survey. They were removed from the sample and ignored for the rest of the study. 
 
\subsection{Photometry \label{sec:photometry}}

To include as much photometric information as possible for each source, we searched the literature and different public surveys for photometric measurements from the optical to the mid-IR.

\subsubsection{Surveys}

We retrieved information from WISE \citep{WISE}, Tycho-2 \citep{Tycho-2}, the Sloan Digital Sky Survey Data Release 8 (SDSS DR8), UKIRT Infrared Deep Sky Survey (UKIDSS), DENIS \citep{Denis}, AKARI \citep{AKARI}, and the Carlsberg Meridian Catalog 14 (CMC 14). To take the different Point Spread Function sizes of each survey  into account, we established an optimized search radius for each of them by fitting a Gaussian to the kernel density distribution of separations between the original (2MASS) and catalog coordinates. The search radius was then set to the average of the Gaussian + 3\,$\sigma$. As a sanity check, we verified that the search radius was consistent with the internal accuracy of the different surveys. When the number of sources was too small to derive a meaningful density distribution of separations, the search radius was set to 1\,\arcsec. This value is a conservative compromise between the astrometric accuracy of the different surveys and the observed separation distribution. During this procedure, upper limits were treated with caution to include them as such. We chose to conservatively redefine as upper limits all fluxes with a signal-to-noise ratio (S/N) $<$ 5 in any 2MASS or WISE bands. We also quadratically added WISE absolute calibration errors of 2.4, 2.8, 4.5 and 5.7\,\% (respectively to its bands) as recommended in the WISE All Sky Catalog Explanatory Supplement.

\subsubsection{Optical to mid-IR photometry from the literature}

We complemented the all-sky optical surveys mentioned above with our own optical ($r$ and/or $i$) photometry from the Canada France Hawaii telescope, when available (Bouy et al., in prep.).

A number of associations presented in Table~\ref{tab:sample} have been observed with the \textit{Spitzer} Space Observatory in the course of various legacy and other programs, and we included the corresponding IRAC and MIPS photometry. The {\it Spitzer} photometry proved to be extremely important for our study, since protoplanetary disks emit most of their light at mid-IR wavelengths. According to the {\it Spitzer} User Manual, absolute calibration of {\it IRAC} and {\it MIPS} fluxes is accurate to $\sim$5\%, and we add the corresponding error quadratically when not already included.

In both cases (optical and mid-IR photometry), a cross-match radius of 1\,\arcsec was used.

\subsection{Resolved binaries and ultra-cool objects}\label{sect:resolved_binaries}

Multiple systems are frequent among stars \citep{Lada2006binaries}, and most of them are not resolved with the ground based or space ({\it Spitzer} and WISE) observations used in this work. The spectral type and photometry of such unresolved sources result in biased measurements. To minimize this problem, we performed an internal match for every region with a search radius of 2\arcsec. A total of 68 objects (34 pairs, $\sim$2\,\% of the whole sample) were found and removed from the sample. Closer systems cannot be identified in the current dataset. We nevertheless note that the properties of multiple systems, and in particular their frequency and distribution of separation, are expected to remain largely unchanged after 1~Myr. Multiplicity frequency is also found to vary little between associations and clusters \citep{Duchene2013} and should equally affect the various groups studied in this work. The statistics (and in particular the disk fractions) derived and discussed in the present study should therefore be regarded as statistics for the systems, rather than for individual sources.

Finally, we also removed objects with spectral types $\leq$ L0. These objects, although very interesting for brown dwarf formation studies, are in most cases beyond the completeness limit of the various surveys used in our analysis. After this process, the sample is comprised of 2\,452 sources.

\section{Data processing}

\subsection{SED cleaning and A$_{\rm V}$ fitting}\label{sect:fitting}

First, the spectral types were converted to T$_{\rm eff}$ using:
\begin{itemize}
\item \citet{Schmidt1982} scale for spectral types K8 and earlier.
\item \citet{Luhman2003} scale for spectral types later than M0.
\end{itemize}

In spite of all the precautions described in Section 2, a blind cross-match of large datasets inevitably contains errors and problems. We therefore visually inspected all the SEDs and rejected problematic sources and photometric measurements. 

Extinction toward each source was then estimated by comparing the observed SEDs with a grid of reddened photospheric BT-Settl models from \citet{Allard2012} of the corresponding T$_{\rm eff}$, and the extinction law of \citet{Indebetouw2005}. Normalization of the observed and synthetic spectra was done at the 2MASS J or H band (in this order or priority) when present, and at the closest band otherwise. The photometric bands used for the $A_{\rm V}$ fitting were:
\begin{itemize}

\item All the optical bands, except those displaying obvious excesses. Some optical bands can indeed be strongly affected by emission lines related to accretion and outflows \citep{Shu1994,Hartmann1994,Gullbring2000}. This is specially the case for the $u$ band, which was systematically ignored in all fits. 

\item All the near-IR bands up to $K_S$, included. Excess in these bands is rarer and usually moderate. We therefore chose to use all of them to maximize the number of measurements used for the fit. Small excesses at these bands result in slightly higher A$_{\rm V}$ values, but are in general tempered by the optical measurements and result mostly in a larger uncertainty.

\end{itemize}
 
Following Bayes theory, the best A$_{\rm V}$ is obtained from the distribution of likelihood among all values computed over the grid. The explored A$_{\rm V}$ range was 0 to 20~mag in steps of 0.1~mag. At least two photometric measurements in the mentioned wavelength range were required for the A$_{\rm V}$ to be fitted.  For 10 sources, only one measurements was available and the A$_{\rm V}$ could not be properly fitted. After this cleaning process, the final sample is comprised of 2\,340 objects.

\begin{figure}
\caption{Comparison between the obtained A$_{\rm V}$ and the values in the literature. The red square shows the typical uncertainty of spectroscopic A$_{\rm V}$ estimates from the literature. The identity relation is represented as a dashed line to guide the eye.}\label{fig:av}

\includegraphics[width=\hsize]{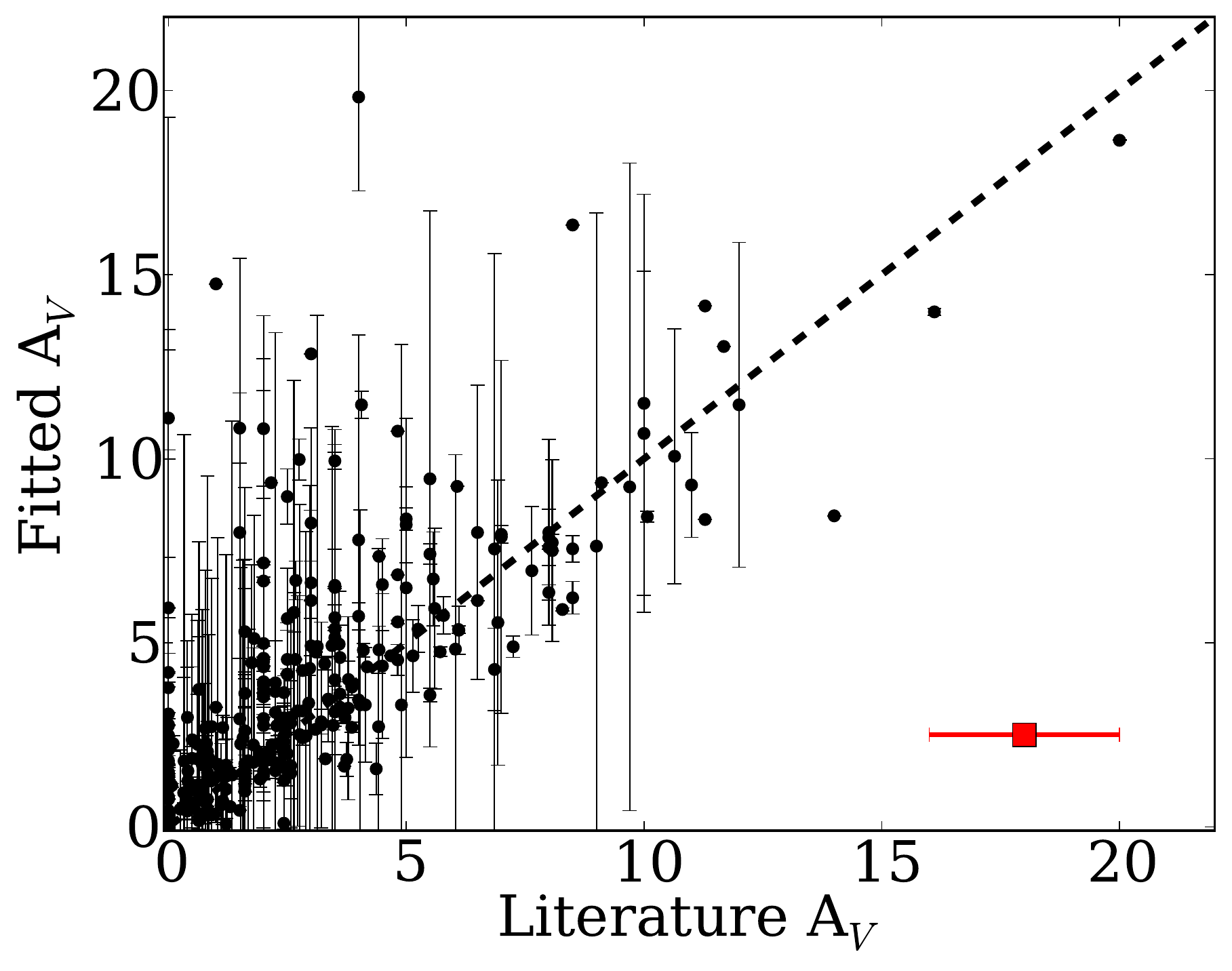} 
\end{figure}

Figure \ref{fig:av} compares the results of the A$_{\rm V}$ fitting process with the values from the literature when available. In general, our measurements are in agreement with previous estimates within the large uncertainties. Note that these estimates and uncertainties are only tentative. Both our measurements and the measurements from the literature are affected by a number of physical and observational biases (variability, accretion, excesses, photometric errors, rotation,etc) which cannot be accounted for. The dispersion in Fig.~\ref{fig:av} gives a more realistic estimate of the real uncertainties on the A$_{\rm V}$.

\subsection{Discrepancy between \textit{Spitzer} and {\it WISE} data}\label{sect:wise_problem}

A discrepancy between \textit{Spitzer} and WISE data was identified during the visual inspection process. We interpret it as the result of contamination from extended background emission and/or source confusion. A more detailed discussion on this issue is given in Appendix A. Data from the AKARI mission (with PSF sizes larger than those of \textit{Spitzer} and WISE) were also discarded for the same reason.

We therefore discarded all the WISE measurements except for nearby associations from the SACY survey. Given their proximity ($<$\,100\,pc), the WISE photometry should not be affected by either extended emission (which is rare in these evolved associations) or confusion (also unlikely, since these associations are extremely lose and extended). Visual inspection of the corresponding SEDs and comparison with {\it Spitzer} photometry when available confirms that the WISE measurements of SACY sources are indeed not affected by any of these issues, but objects flagged as extended in the WISE catalog were conservatively discarded anyway.

\section{Disk fractions}\label{sect:identifying}

In this section we study disk fractions as probed by mid-IR excesses. For the rest of this work, any object with excess at any wavelength longer than 3\,$\mu$m is considered as a disk-bearing system. Prior to identifying disks and deriving disk percentages, we investigated the sensitivity and spatial completeness of the various mid-IR surveys.

\subsection{Sensitivity completeness}\label{sect:sensitivity}

Circumstellar disks produce a flux excess at IR wavelengths, and hence disk-bearing stars are easier to detect than pure photospheres. A magnitude limited sample such as the one presented here can thus be biased toward objects with disks, leading to overestimated disk fractions. We corrected from this bias by discarding objects with photospheric flux values (as obtained from models used during the SED fitting, see Sect.~\ref{sect:fitting}) below the detection limits of the different instruments and observations. According to the user manual, the WISE mission (used for the SACY sample) surveyed the whole sky with 5-$\sigma$ sensitivity limits of 0.08, 0.11, 1 and 6~mJy for the W1, W2, W3, and W4 bands respectively. We estimate that the W1, W2 and W3 sensitivities encompass the entire luminosity range of the SACY sample, while the W4 band is sensitive down to the late-K/late-M dwarfs, depending on the age and distance.
\textit{Spitzer} data (used for the rest of the associations, see Table~\ref{tab:sample}) come from various surveys with different sensitivities, summarized in Table~\ref{tab:sensitivity}. While the optical, near-IR and {\it IRAC} data are sensitive to the entire luminosity range described in Section~\ref{sect:mass_coverage}, the shallower {\it MIPS} observations define a more limited completeness domain. We estimate that the {\it MIPS} sensitivity described in Table~\ref{tab:sensitivity} corresponds to early-M/early-K dwarfs depending on the age and distance, respectively. 

\begin{table*}
  \caption{Sensitivity completeness limits for the \textit{Spitzer} photometric data}
  \label{tab:sensitivity}
  \begin{center}
    \begin{tabular}{lccccccccc}
      \hline\hline Regions & IRAC1 (mJy) & IRAC2 (mJy) & IRAC3 (mJy) & IRAC4 (mJy) & MIPS1 (mJy)\\
\hline
      All regions (except Taurus) & 0.08 & 0.05 & 0.12 & 0.16 & 1.03\\
      Taurus \citep{Rebull2011} & 0.18 &  0.12 &  0.62 &  0.62 &  5.12\\
      \hline
    \end{tabular}
  \end{center}
\end{table*}

\subsection{Spatial coverage}
WISE is an all-sky survey, and hence the SACY regions -- for which only WISE data was used -- were fully covered in the mid-IR. On the other hand, {\it Spitzer} observations used in this study are covering only fractions of the various associations, resulting in a number of non detections. These non detections were naturally rejected in the previous step.

\subsection{Establishing the presence of disks}\label{sect:disk_criterion}

Circumstellar disks produce mid/far IR excess in relation to photospheres \citep{Williams2011}. Dust around stars is heated by stellar radiation and re-radiates with typical T$_{\rm eff}$ of a few tens or hundreds of kelvin. The amount of excess as a function of wavelength depends on the disk properties, resulting in a wide diversity of SED shapes and making the detection of circumstellar disks sometimes difficult using a single color or luminosity.

To identify excesses, we use the significance index $\chi_{\lambda}$, defined as $\chi_{\lambda} ={\rm (F_{observed, \lambda } - F_{model, \lambda })/\sigma_{{\rm Observed},\lambda}}$, where ${\rm F_{observed, \lambda }}$ is the observed flux, ${\rm F_{model, \lambda }}$ is the corresponding model photospheric flux, and $\sigma_{{\rm Observed},\lambda}$ is the error at the corresponding wavelength ($\lambda$). We define as ``excess'' any measurement displaying a $\chi_{\lambda} \geq 5$, corresponding to a 5-$\sigma$  detection over the expected photospheric value. We note that flux uncertainties play an important role in the definition of $\chi$ and were therefore considered with caution (see Section~\ref{sec:photometry}). Subestimated errors would indeed result in artificially high disk percentages, while over-estimated error would result in artificially low disk fractions. 

To investigate the dependence of the presence of excess as a function of wavelength, we defined three wavelength regimes:
\begin{itemize}
\item Short wavelength range: sources displaying excess in any band between 3.4--4.6~$\mu$m (including IRAC1, IRAC2, WISE1, WISE2),
\item Intermediate wavelength range: sources displaying excess in any band between 8.0--12.0~$\mu$m (including IRAC4, WISE3),
\item Long wavelength range: sources displaying excess in any band between 22--24~$\mu$m (including MIPS1 and WISE4).
\end{itemize} 

The mid-IR emission at different wavelengths probes the disk at different distances from the host star and depth within the disk. Table~\ref{tab:criteria} gives an overview of the spatial domains probed by the three wavelength ranges defined above. These values, based on simplifications, should be regarded as indicative only. They nevertheless provide reasonable estimates of the origin of most of the corresponding emission.

Mid-IR excess related to the presence of disks have often been identified using color-color and color-magnitude diagrams. To compare our results with these studies, we also computed fractions for objects detected in all IRAC bands and displaying excess in at least one of them (all regions but SACY) or objects detected in all W1, W2, and W3 bands and excess in any of these three bands (SACY associations). 
\begin{table*}
  \caption{Region of the disk probed by the three wavelengths ranges. }
  \label{tab:criteria}
  \begin{center}
    \begin{tabular}{c c c}
      \hline\hline 
      Wavelength range & Disk radius & Label\\
      ($\mu m$)  & (AU) \\
      \hline 
      3.4-4.6  & 0.01 - 1 & Short \\
      8-12 &  0.03- 5 & Intermediate \\
      22-24 & 0.3 - 60 & Long \\
\hline
\end{tabular}
    \end{center}
\end{table*}

The ``excess'' definition based on a conservative 5-$\sigma$ level of significance might not be optimized for the detection of debris disks. These evolved disks typically display significantly weaker excesses compared to protoplanetary disks \citep{Wyatt2008}. The corresponding debris disk fractions should therefore be regarded as estimates within the boundaries set by the definition of ``excess'' used here, rather than absolute values. The same applies for near-IR excesses, these being intrinsically weaker than mid-IR ones. Our choice of a more conservative threshold (5 instead of the``usual'' 3\,$\sigma$) is motivated by the problems encountered with the WISE photometry (see Sect.~\ref{sect:wise_problem} and Appendix), which is used mainly for the older regions and therefore for the objects with the weaker mid-IR excesses. We also note that the usage of two photometric bands in each of the considered wavelength ranges (short, intermediate, long) should mitigate this possible bias.

\subsection{Disk fraction's values and errors}

Disk fractions were calculated as the ratio of objects displaying an excess (as defined before) over the total number of sources after completeness corrections. The large size of our sample allowed us to perform this analysis for the three wavelength ranges defined previously.

Estimating uncertainties on disk fractions using standard error propagation is impractical, as uncertainties on the photometry (errors related to measurement, but also variability, rotation, activity, etc), the fit (A$_{\rm V}$) and various models (adopted extinction law, Spectral Type vs $\rm T_{eff}$ relation) are difficult to assess. We chose to estimate errors in an empirical way by using bootstrapping (1000 iterations for each region) and randomly varying the T$_{\rm eff}$ and observed fluxes within normal distributions (with $\sigma$ of 100\,K and $\sigma_{\rm flux}$, respectively). For regions with a high number of sources  ($>$\,30), disk fraction errors are set as the standard deviation obtained with the boostrapping process. For smaller samples, the errors were corrected using the corresponding Student's t distribution with a confidence interval of 68\,\% (comparable to the 1\,$\sigma$ value for larger samples). The derived disk fraction values are given in Table~\ref{tab:fractions} and Figs.~\ref{fig:disk_fractions} and~\ref{fig:prim_evolv}. Because of the different sensitivity and spatial completeness levels of the sample, the number of sources varies in each bands. To avoid problems related to small number statistics, we discarded datasets with less than ten objects per region and wavelength range.

\begin{table*}
  \caption{Fraction of sources with excess for each region in the three wavelength regimes.}
  \label{tab:fractions}
  \begin{center}
    \begin{tabular}{lcccc}
      \hline\hline Name & Fraction $_{\rm IRAC}$(\%) & Fraction $_{\rm short}$(\%) & Fraction $_{\rm intermediate}$(\%) & Fraction $_{\rm long}$(\%) \\
      \hline
      25~Orionis & 9 $\pm$ 5 [33] & 6 $\pm$ 4 [34] & 9 $\pm$ 5 [33] & \ldots [0]\\
      Cha~I  &  52 $\pm$ 6 [86] &  43 $\pm$ 5 [109] & 51 $\pm$ 5 [102] & 66 $\pm$ 9 [29]\\
      Cha~II &  84 $\pm$ 9 [19] & 61 $\pm$ 9 [28] & 84 $\pm$ 9 [19] & 87 $\pm$ 9 [15]\\
      CrA     &  50 $\pm$ 13 [16]  &  31 $\pm$ 11 [19] & 50 $\pm$ 13 [16] & \ldots [4]\\
      IC~348  &    36 $\pm$ 3 [240] &  29 $\pm$ 3 [262] & 30 $\pm$ 3 [240] & 50 $\pm$ 17 [10] \\
      $\lambda$-Orionis  & 26 $\pm$ 7  [43] & 20 $\pm$ 6 [51] & 26 $\pm$ 6 [43] & \ldots [0] \\
      Lupus  &   52 $\pm$ 5 [85]  & 38 $\pm$ 5 [89] & 52 $\pm$ 5 [87] & 43 $\pm$ 7 [49] \\
      NGC~1333  &   66 $\pm$ 6  [67] &  47 $\pm$ 6 [73] & 66 $\pm$ 6 [67] & \ldots [2]\\
      Ophiuchus  &   25 $\pm$ 3 [214] & 17 $\pm$ 3 [253] & 25 $\pm$ 3 [228] & 51 $\pm$ 8 [41] \\
      $\sigma$-Orionis  &  39 $\pm$ 6  [70] &  18 $\pm$ 5 [74] & 39 $\pm$ 6 [70] & \ldots [0]\\
      Serpens &  62 $\pm$ 4 [125]  &  44 $\pm$ 4 [131] & 60 $\pm$ 5 [125] & 32 $\pm$ 11 [29]\\
      Taurus &  63 $\pm$ 4  [180] &  48 $\pm$ 3 [212] & 63 $\pm$ 4 [183] & 58 $\pm$ 14 [12] \\
      Upper Sco &  16 $\pm$ 6 [32] &  7 $\pm$ 2 [250] & 11 $\pm$ 2 [232] & 22 $\pm$ 3 [165] \\
      \hline
      AB~Dor  & 5 $\pm$ 3 [78] & 4 $\pm$ 3 [78] & 3 $\pm$ 2 [77] & 4 $\pm$ 3 [51] \\
      Argus    &   4 $\pm$ 5 [44]  & 0 $\pm$ 3 [44] & 5 $\pm$ 5 [44] & \ldots [3]  \\
      $\beta$ Pic   &  18 $\pm$ 6 [50] &  16 $\pm$ 5 [50] & 8 $\pm$ 4 [59] &  22 $\pm$ 6 [46]\\
      Carina     &   4 $\pm$ 6 [27]  &  0 $\pm$ 3 [27] & 4 $\pm$ 5 [28] &  \ldots [6] \\
      Columba   &   5 $\pm$ 4  [57] &  2 $\pm$ 2 [57] & 5 $\pm$ 4 [57] & 28 $\pm$ 10 [18] \\
      $\eta$ Cha  &  21 $\pm$ 9 [24]  &  17 $\pm$ 8 [24] & 21 $\pm$ 9 [24] & 30 $\pm$ 15 [10]\\
      Octantis    &   6 $\pm$ 7  [17]  &  0 $\pm$ 4 [17] & 6 $\pm$ 7 [17] & \ldots [1] \\
      Tuc-Hor    &     8 $\pm$ 4 [48]  &  8 $\pm$ 4 [48] & 0 $\pm$ 3 [48] & 12 $\pm$ 5 [41] \\
      TW~Hya       &   20 $\pm$ 9 [19] &  5 $\pm$ 6 [19] & 16 $\pm$ 8 [19] & 29 $\pm$ 12 [14]\\
      \hline
    \end{tabular}
\tablefoot{The number of sources used is indicated within brackets. Subindices reference fractions at different wavelength ranges, as defined in Sec.~\ref{sect:disk_criterion}}
  \end{center}
\end{table*}

\section{Disk fractions as a function of age}\label{sect:disk_frequencies}

\begin{figure*}
\caption{Upper left panel: comparison of the disk fractions obtained in this study (circles) with those from  \citep{Hernandez2007,Hernandez2008} (squares) for regions in common.  Upper right and lower panels: disk fractions for all regions in the short, intermediate and long wavelength ranges. The best fit exponential law is over-plotted (red line). }\label{fig:disk_fractions}
\includegraphics[width=0.95\textwidth]{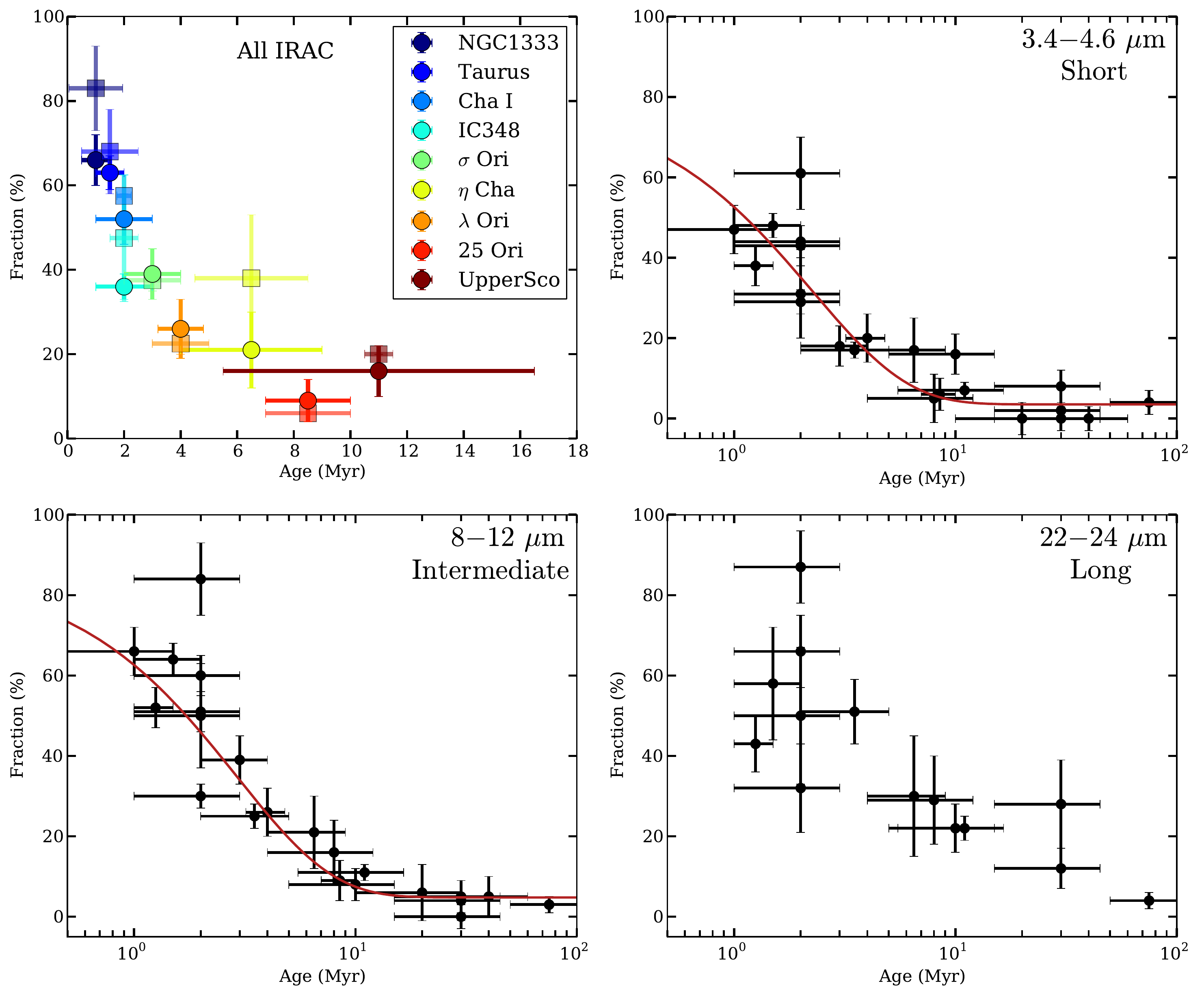}
\end{figure*}

In this section, we investigate disk dissipation timescales as a function of age and wavelength. As a sanity check, we first compare our results with previous studies from \citet{Hernandez2007} and \citet{Hernandez2008} for the nine star-forming regions in common: 25~Orionis, Cha~I, $\eta$-Cha, IC~348, $\lambda$-Orionis, NGC~1333, $\sigma$-Orionis, Taurus and Upper Sco. Figure~\ref{fig:disk_fractions} shows that the results are in good agreement within the estimated uncertainties. We also compare our results for part of the SACY sample from those in \citet{Zuckerman2011}: all but one (AB Dor) associations are within a 2\,$\sigma$ range. We note that our values are almost systematically lower than the ones given in the literature, as expected given the more conservative disk detection criterion used in this work (see Sec.~\ref{sect:disk_criterion}). 

We then study the excess fraction for each of the three wavelength ranges as a function of age (Fig.~\ref{fig:disk_fractions}, top right and lower panels). Our results show a decay of disk fractions with age at all wavelength ranges The values obtained for Cha~II are considerably higher than those of other associations of similar age, but consistent with the value reported by \citet{Alcala2008}. The current dataset does not allow us to explain this difference, and the cause is outside the scope of this study. Figure~\ref{fig:disk_fractions} shows that disk fractions for the intermediate (8--12 $\mu$m) wavelength range are higher than for the short (3.4-4.6\,$\mu$m) wavelength range. Although of low statistical significance (within 1-$\sigma$), the systematic offset for the entire sample suggests that this relative difference is a real trend.

In this study, we follow \citet{Mamajek2009} and model these distributions with an exponential law of the form:
\begin{equation}
A e^{-t/\tau} + C
\end{equation}
where $t$ is the age of the region (in Myr), and $A$, $\tau$ and $C$ are left as free parameters. In this simple parametric model, $\tau$ can be interpreted as the characteristic timescale of IR excess decay, and has been previously found to have values between 2$\sim$3~Myr \citep{Mamajek2009,Williams2011}. The best fits obtained for $\tau$ in the short (3.4--4.6~$\mu$m) and intermediate (8-12\,$\mu$m) wavelength ranges are 2.3$\pm$0.5~Myr and 2.8$\pm$0.8~Myr respectively, indistinguishable within the uncertainties and in good agreement with the results from \citet{Mamajek2009}. 
The exponential fit has the advantage of modeling the entire time domain with a single function, and was therefore preferred over the linear fit model used in \citet{Haisch2001}. Further investigation of the possible analytic laws to model this behavior and their corresponding physical interpretation are outside the scope of our work.

 
Whereas IR excesses in the short and intermediate wavelength ranges are thought to come from primordial optically-thick disks \citep{Williams2011}, excesses in the long wavelength range (22-24~$\mu$m) can be produced by either primordial, transitional or debris disks \citep{Wyatt2008}.  These different types of disk do not necessarily have the same evolutionary mechanisms and timescales \citep{Wyatt2008}. Fitting a simple exponential law to the long wavelength dataset would therefore have no physical meaning and was not attempted here.

\begin{figure}
\caption{Fraction of primordial (red circles) and evolved (blue squares) disk as a function of age. The best fit exponential law for the primordial disk percentages is overplotted as a red line. The red shaded area represents the fit obtained for two extreme cases: the highest and lowest values at every age.}\label{fig:prim_evolv}
\includegraphics[width=\hsize]{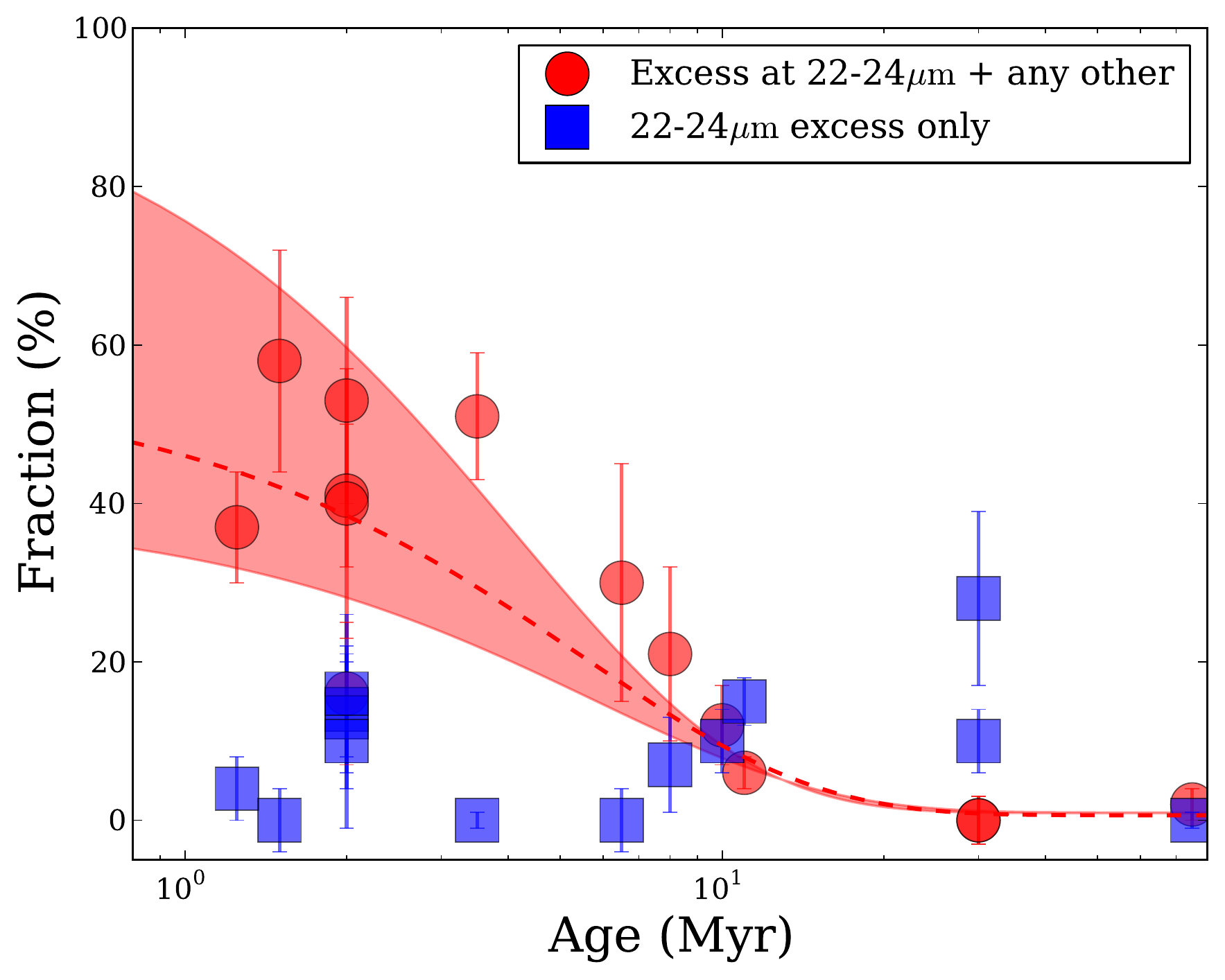} 
\end{figure}

To disentangle disks populations in the long wavelength dataset, we split the sample in two groups:
\begin{itemize}
\item ``Primordial'' disks: sources with excess at 22-24\,$\mu$m \emph{and} at least one shorter wavelength.
\item ``Evolved'' disks: sources with excess at 22-24\,$\mu$m \emph{only}, including mostly transitional and debris disks. In some cases, edge-on disks can mimic the SED of transitional disks and contaminate the group. Their frequency is nevertheless expected to be very low.
\end{itemize}
Fig.~\ref{fig:prim_evolv} shows the fractions of primordial and evolved disks as a function of time. The distribution is more dispersed than those found at shorter wavelengths (Fig.~\ref{fig:disk_fractions}), specially at very young ages ($<$2~Myr). After 2$\sim$3~Myr, it shows a clear and smooth decay and reaches a $\sim$ zero level around 10$\sim$30\,Myr, as observed at shorter wavelengths. To estimate the timescale of excess decay, we fit an exponential law to:
\begin{enumerate}
\item The whole primordial disks dataset: $\tau$=5.5 $\pm$ 1.4~Myr.
\item The highest values at every given age: $\tau$=4.2 $\pm$~0.6~Myr.
\item The lowest value at every given age: $\tau$=5.8 $\pm$ 1.5~Myr.
\end{enumerate}

In all three cases -- including the extremes -- $\tau$ is significantly higher than in the short and intermediate wavelength ranges.  Fitted values with their estimated errors for the short, intermediate, and ``primordial'' long wavelength regimes are given in Table~\ref{tab:fitting_results}.

\begin{table*}
  \caption{Values and errors obtained from fitting and exponential decay to disk fractions with time.}
  \label{tab:fitting_results}
  \begin{center}
    \begin{tabular}{l c c c }
      \hline\hline 
Wavelength range & A & $\tau$ & C  \\
& (\%/Myr) & (Myr) & (\%) \\ 
\hline
Short & 80 $\pm$ 10 & 2.3 $\pm$ 0.5 & 4 $\pm$ 2\\
Intermediate & 80 $\pm$ 10 & 2.8 $\pm$ 0.8 & 5 $\pm$ 2\\
Long (``primordial'') & 40 $-$ 100 & 4.2 $-$ 5.8 & 0 $-$ 2\\
\hline
\end{tabular}
    \end{center}
\tablefoot{In the case of long wavelength range, only ``primordial'' disks were fitted, and hence no value is provided for ``evolved'' disks (see Sect.~\ref{sect:disk_frequencies}).}
\end{table*}

\begin{figure}
\caption{Ratios of the observed and predicted photospheric values in the long wavelength range (22--24\,$\mu$m) as a function of age.}\label{fig:proto_vs_debris}
\includegraphics[width=\hsize]{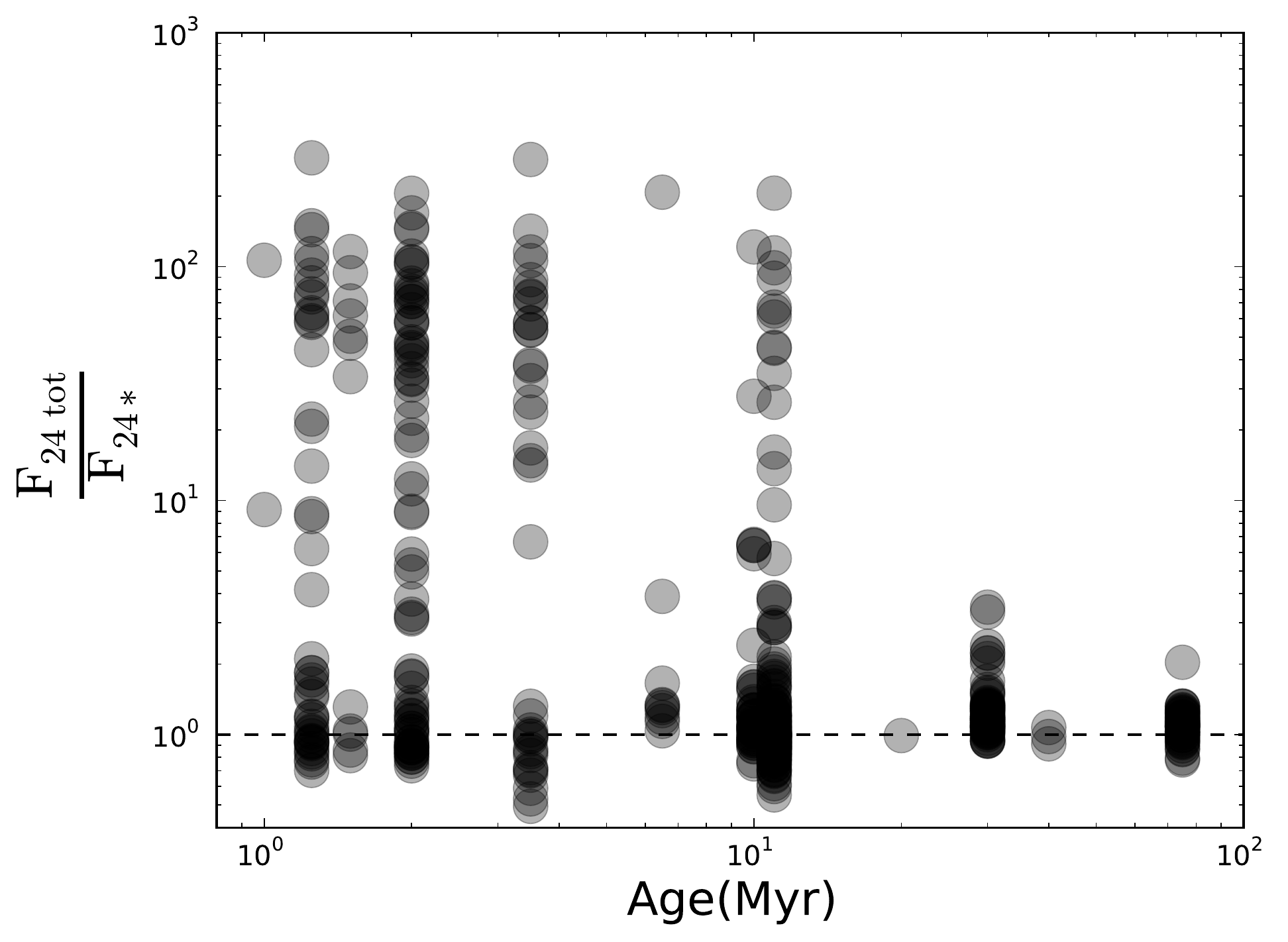}
\end{figure}

\section{Discussion}\label{sect:discussion}

\subsection{Wavelength-dependent disk fractions}\label{wavdiskfreq}

The unprecedented size of our sample and the consistency of our analysis allows for the first time to perform statistically robust measurements of disk fractions as a function of both time and wavelength. Figure~\ref{fig:disk_fractions} shows that below 10~Myr, disk fractions derived from mid-IR excess systematically increase with wavelength  (as shown in Table \ref{tab:fractions}). We note that this difference could be affected by the stringent ``excess'' criterion applied here, which could favor the detection of strong excesses typical of mid-IR bands. We verify that the trend of increasing disk fractions with wavelength up to 10 Myr is still present when a $\chi >$3 cut is used. This result suggests that great care should be taken when comparing studies of disk fractions derived using different wavelength ranges.

\subsection{Disk decay and dust evolution}\label{decay}

The timescale of mid-IR excess decay obtained in the short and intermediate wavelength ranges are found to be similar, suggesting that the corresponding dust populations evolve on a similar timescale. One the other hand, Fig.~\ref{fig:disk_fractions} shows that there is more excess at intermediate wavelengths at a given age than at short wavelengths. The two results combined suggest that the dust emitting at shorter wavelengths -- which is in average closer to the star -- starts evolving sooner.

The higher mid-IR excess fractions at all ages and the longer decay times scale in the long wavelength range (22-24~$\mu$m) both suggest that the dust at larger radii evolves slower. Although marginal, the higher dispersion of the excesses distribution at long wavelengths points toward a less coherent evolution of the dust in that regime than at shorter wavelengths. The dust populations at larger radii are indeed believed to experience both vertical (settling) and radial (mixing, migration, etc) transfers during the disk evolution. Grain growth, leading eventually to planetary embryos, is also affecting the disk properties and could participate to the observed scatter in time.

\subsection{Transition from protoplanetary to debris disks}

Figure~\ref{fig:proto_vs_debris} shows flux ratios (observed over expected photospheric values) in the long wavelength range (22-24\,$\mu$m) for all objects as a function of age. Primordial disks display stronger excesses over the photosphere than their evolved counterparts. A sharp decline in flux excess is observed around 10~Myr, indicating a significant change in the disk properties at that stage. On the other hand, Figs.~\ref{fig:disk_fractions} and \ref{fig:prim_evolv} show that disk fractions reach a $\sim$ zero level at about the same time. These two results confirm the disk lifetime of $\approx$10~Myr reported previously in the literature \citep{Haisch2001,Hernandez2007,Hernandez2008,Mamajek2009,Williams2011}. Unfortunately, current data do not provide information on whether the material found in young debris disks are remnants of the protoplanetary stage, or it is a different dust population produced by planetesimal collisions \citep{Wyatt2008}. Dissecting this transition from the gas rich to gas poor phases is one of the overall long-term goals of this project.

Tendencies for ``evolved'' disks are much harder to determine given their low number at all ages. However, two interesting features can be identified in Fig.~\ref{fig:prim_evolv}. First, ``evolved'' disks are not present (or extremely uncommon) at young ages ($\lesssim$ 5 Myr).  Second, the fraction of ``evolved'' disk seems to show a bump starting at $\sim$8\,Myr, matching the disappearance of the ``primordial'' disks. The latter feature, which needs to be confirmed with a larger sample, reproduces the results by \citet{Currie2008} who reported an increase of mid-IR emission from debris disks at 5--10\,Myr, followed by a peak at 10--15\,Myr, and a decrease from there on in the double cluster h and $\chi$ Persei. They argue that this qualitatively matches the predictions from the debris disk evolution tracks by \citet{Kenyon2004b} for disk masses between 3 and 1/3 of the Minimum Mass Solar Nebula, and that kilometer-sized planetesimals already formed between 0 and 2~Myr. This suggest a disk self-stirring process \citep{Wyatt2008}, which would fill the system with dust produced by collisions of planetesimals and increase the mid-IR flux for some Myr at 22-24\,$\mu$m. As a result, a higher debris disk detection rate will be obtained, and hence higher disk numbers. The combination of these two features (no ``evolved'' disks at young ages, plus their fractions tentatively increasing when ``primordial'' disks start disappearing) reinforces the idea of ``evolved'' disks being the outcome of ``primordial'' disk evolution. Interestingly, the disappearance of ``primordial'' disks, and hence gas \citep{Sicilia-Aguilar2005,Sicilia-Aguilar2006}, seems to match in time the period of dust stirring, which indeed require negligible gas to be effective \citep{Wyatt2008}.

\section{Conclusions}\label{conclusions}

We have compiled a large sample of 2\,340 spectroscopically confirmed young stellar objects from 22 nearby ($< 500$\,pc) associations. Spectral energy distributions covering from visible to mid-IR (22-24~$\mu$m) wavelengths have been built and visually inspected for all the sources.  We studied the disk fraction as probed by excess over the photosphere in three wavelength ranges: 3.4--4.6~$\mu$m (short) , 8--12~$\mu$m (intermediate) and 22--22~$\mu$m (long), and find:

\begin{itemize}
\item The excess fraction is dependent on the wavelength range considered. Transforming excess percentages to disk percentages is therefore not straightforward, and special care must be taken when comparing disk fractions derived using differing wavelengths.
\item Significant discrepancies between the \textit{Spitzer} and WISE photometry are found, mostly due to confusion and extended background emission, and calling for a revision of previous studies based on blind analysis of the WISE luminosities. 
\item Mid-IR excesses are systematically more frequent at longer wavelength, compatible with inside-out disk clearing scenarios.
\item For primordial disks, the dust probed between 22--24\,$\mu$m seems to evolve more slowly. Assuming an exponential decay, we derive a timescale $\tau$ of 4$\sim$6~Myr at 22--24~$\mu$m, compared to 2$\sim$3~Myr at shorter wavelengths. This result is also compatible with inside-out disk clearing.
\item There is more dispersion in the fraction of excess sources with age when measured at 22--24~$\mu$m in comparison to shorter wavelengths, suggesting that other mechanisms are involved in the evolution of the dust at larger radii.
\item The fraction of ``evolved'' disks seems to rise around 8-10~Myr, and decrease again afterward.
\item The disappearance of ``primordial'' (potentially gas-rich) disks at 10$\sim$20~Myr  matches in time the brief rise of the number of ``evolved'' disks. This result is compatible with planet formation theories suggesting that the disappearance of the gas is immediately followed by the dynamical stirring of the debris disk produced by planetesimal collisions that will ultimately lead to planet formation.
\item These last two conclusions reinforce the idea of ``evolved'' disks being the outcome of ``primordial'' disk evolution.
\end{itemize}

Future studies based on the present sample will extend these results, obtained for the integrated stellar mass spectrum, by investigating disk fractions and properties as a function of stellar mass, age, and environment. The inclusion of far-IR data from \textit{Herschel} will ultimately provide crucial information about outer disk regions and disk masses, contributing to a much more complete picture of disk evolution.

\begin{acknowledgements}
This work has been possible thanks to the ESAC Science Operations Division research funds with code SC 1300016149, support from the ESAC Space Science Faculty and of the Herschel Science Centre. We thank the referee for valuable comments that helped to improve the paper, as well as Eric Mamajek for additional suggestions. We also want to thank Neal J. Evans II, Ewine F. van Dishoeck, Paul Harvey, Geoffrey A. Blake, Fernando Comer\'on, Lucas Cieza, Isa Oliveira, Luisa Rebull, Juan Alcal\'a, Joanna Brown, Jean-Charles Augereau, Vincent Geers and other members of the c2d team for comments to a previous c2d works on the same topic that had to be abandoned due to inappropriate data for the intended study. H. Bouy is funded by the Spanish Ram\'on y Cajal fellowship program number RYC-2009-04497. This work has made an extensive use of Topcat \citep[\url{http://www.star.bristol.ac.uk/~mbt/topcat/},][]{2005ASPC..347...29T}. This work is based in part on data obtained as part of the UKIRT Infrared Deep Sky Survey. This research made use of the SDSS-III catalog. Funding for SDSS-III has been provided by the Alfred P. Sloan Foundation, the Participating Institutions, the National Science Foundation, and the U.S. Department of Energy Office of Science. The SDSS-III web site is http://www.sdss3.org/. SDSS-III is managed by the Astrophysical Research Consortium for the Participating Institutions of the SDSS-III Collaboration including the University of Arizona, the Brazilian Participation Group, Brookhaven National Laboratory, University of Cambridge, University of Florida, the French Participation Group, the German Participation Group, the Instituto de Astrofisica de Canarias, the Michigan State/Notre Dame/JINA Participation Group, Johns Hopkins University, Lawrence Berkeley National Laboratory, Max Planck Institute for Astrophysics, New Mexico State University, New York University, Ohio State University, Pennsylvania State University,  University of Portsmouth, Princeton University, the Spanish Participation Group, University of Tokyo, University of Utah, Vanderbilt University, University of Virginia, University of Washington, and Yale University. 
This work makes use of data from the DENIS Survey. DENIS is the result of a joint effort involving human and financial contributions of several Institutes mostly located in Europe. It has been supported financially mainly by the French Institut National des Sciences de l'Univers, CNRS, and French Education Ministry, the European Southern Observatory, the State of Baden-Wuerttemberg, and the European Commission under networks of the SCIENCE and Human Capital and Mobility programs, the Landessternwarte, Heidelberg and Institut d'Astrophysique de Paris. 
This publication makes use of data products from the Wide-field Infrared Survey Explorer, which is a joint project of the University of California, Los Angeles, and the Jet Propulsion Laboratory/California Institute of Technology, funded by the National Aeronautics and Space Administration.

This research used the facilities of the Canadian Astronomy Data Centre operated by the National Research Council of Canada with the support of the Canadian Space Agency.   

This publication makes use of data products from the Two Micron All Sky Survey, which is a joint project of the University of Massachusetts and the Infrared Processing and Analysis Center/California Institute of Technology, funded by the National Aeronautics and Space Administration and the National Science Foundation.

This work is based in part on observations made with the Spitzer Space Telescope, which is operated by the Jet Propulsion Laboratory, California Institute of Technology under a contract with NASA.

\end{acknowledgements}

\bibliographystyle{aa}
\bibliography{biblio2}

\begin{appendix}

\section{On the discrepancy between \textit{Spitzer} and WISE data}

\begin{figure*}
\caption{Differences between WISE and \textit{Spitzer} data as a function of \textit{Spitzer} fluxes for the three comparable bands. Only WISE detections with S/N > 5 are plotted in each band.}\label{fig:spitzer_wise_problem}
\includegraphics[width=\hsize]{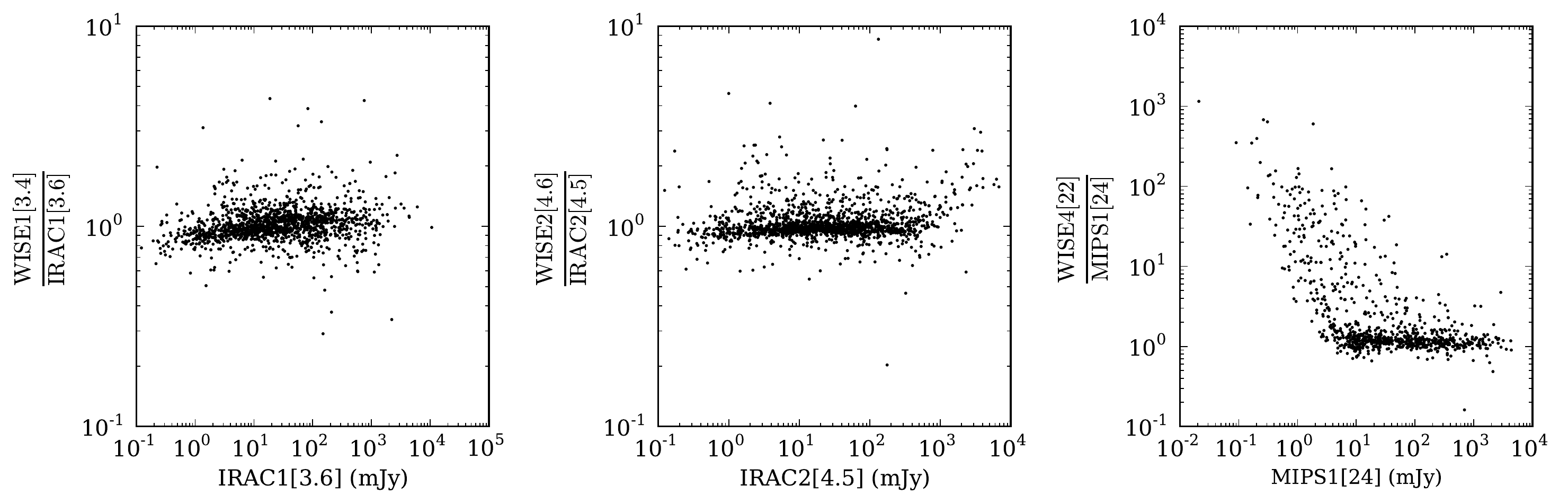}
\end{figure*}

\begin{figure*}
\caption{Three-color composite images of IC~348 as seen by \textit{Spitzer} (left, blue: IRAC1, green: IRAC2, and red: MIPS1) and WISE (right, blue: W1, green: W2, and red: W4). The blue dots marks the position of 2MASSJ03443274+3208374, which SED is shown in Fig.~\ref{fig:sed_wise_spitzer} \label{fig:ic348}}
\includegraphics[width=0.49\hsize]{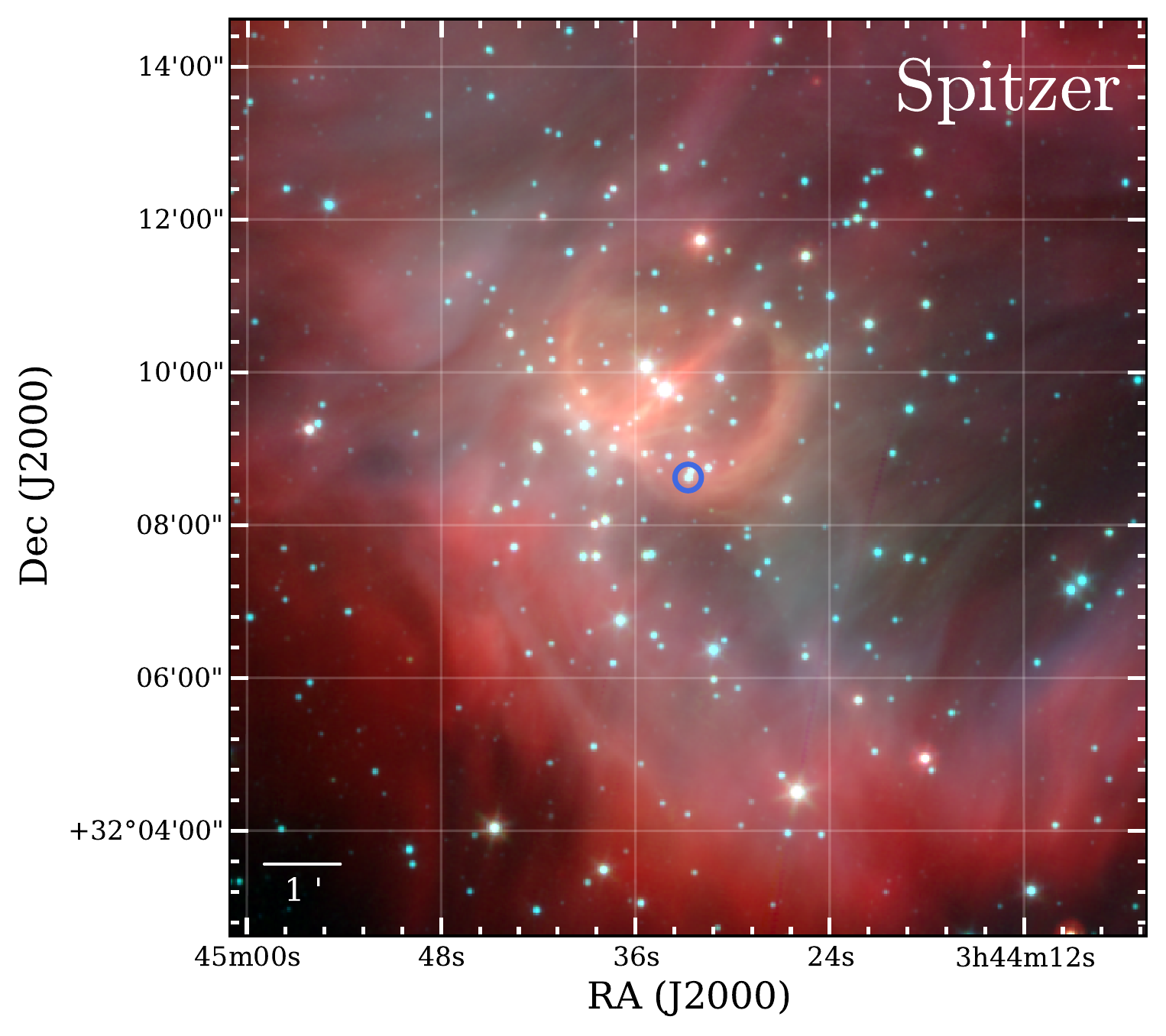}  \includegraphics[width=0.49\hsize]{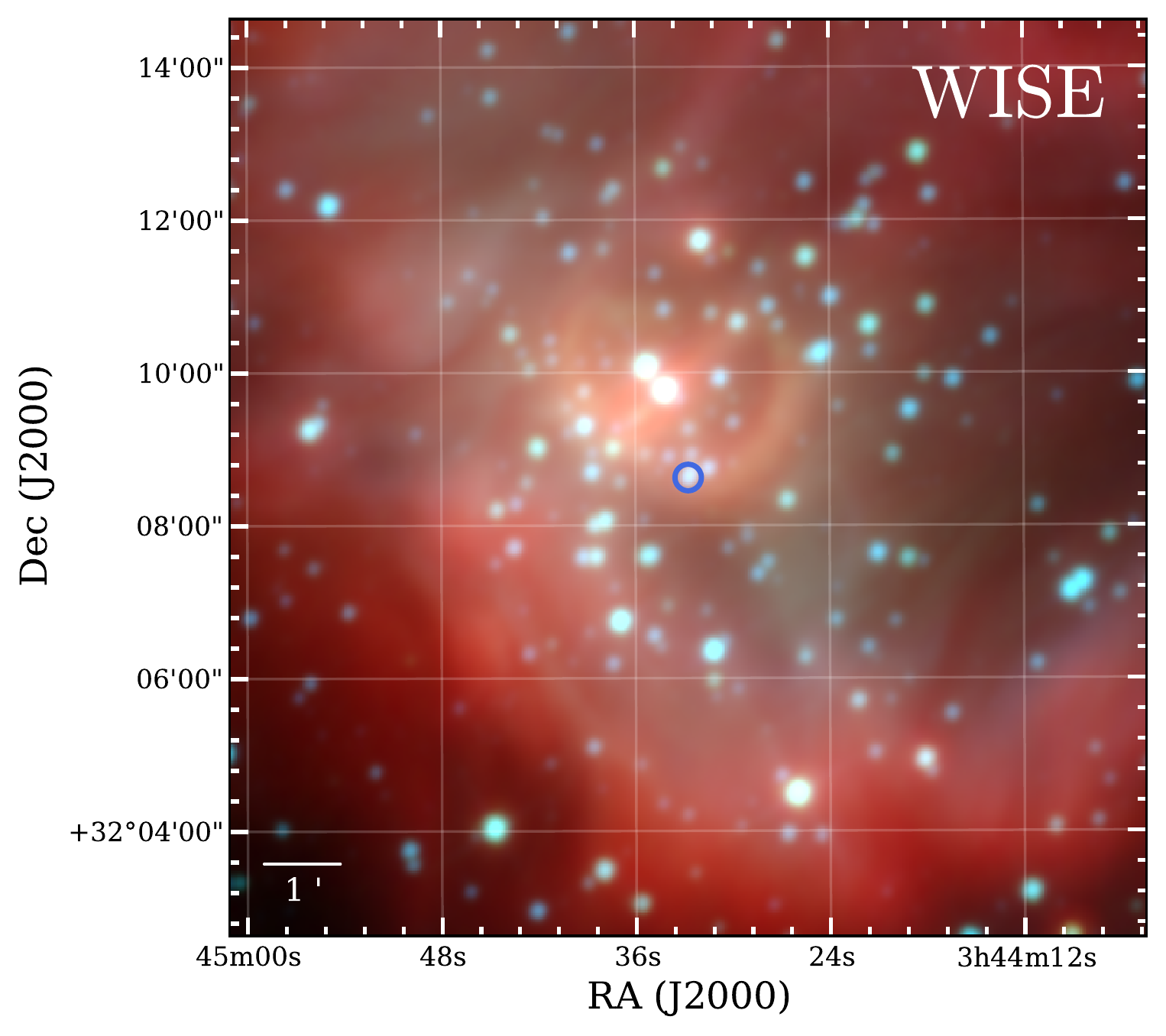} 
\end{figure*}

\begin{figure}
\caption{SED of 2MASSJ03443274+3208374, a source located in a region of high density and variable extended emission (see Fig.~\ref{fig:ic348}). \label{fig:sed_wise_spitzer}}
\includegraphics[width=0.49\textwidth]{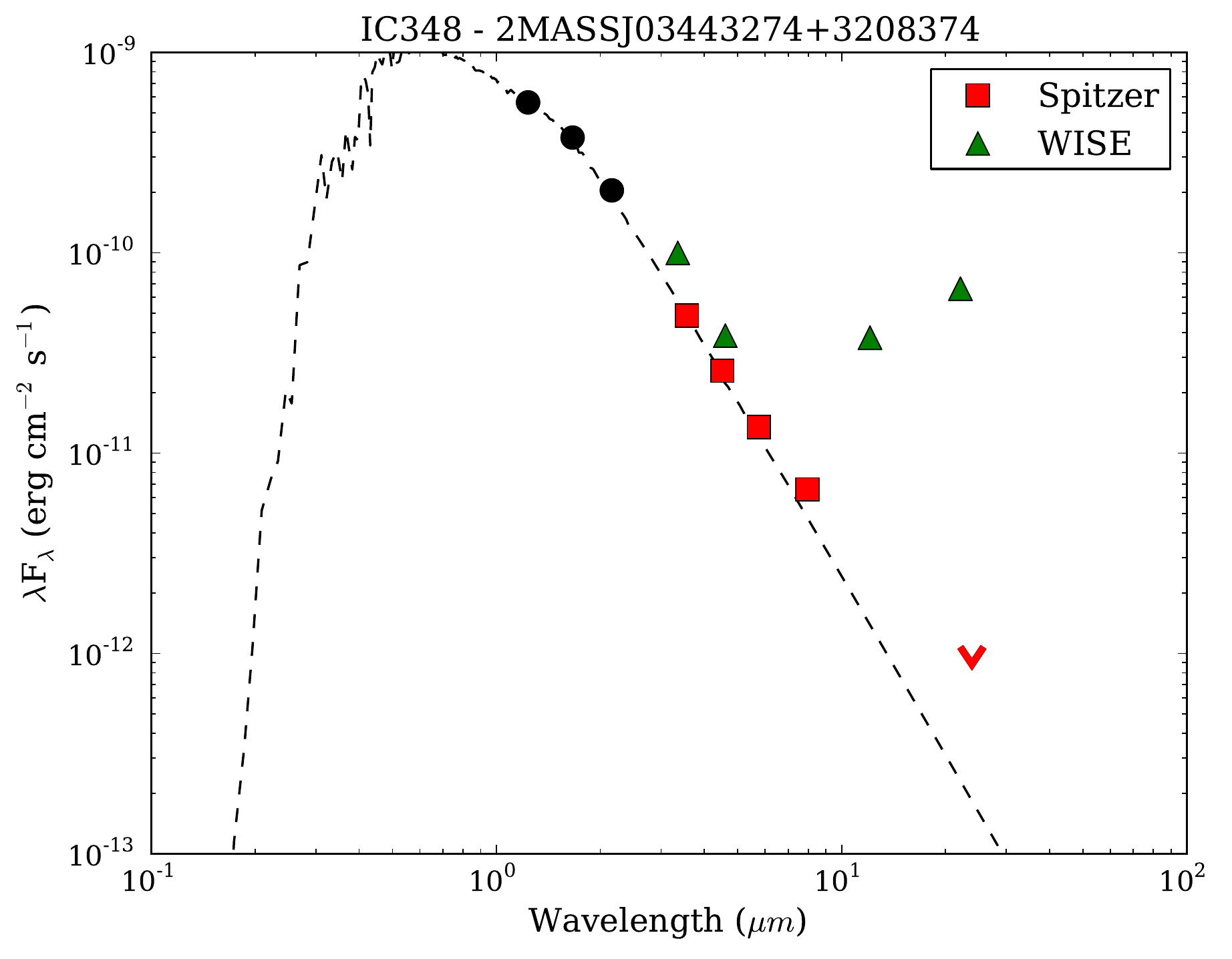} 
\end{figure}

The WISE mission used four IR bands (3.4, 4.6, 12, and 22\,$\mu$m, W1, W2, W3 and W4 respectively) to survey the whole sky. On the other hand, the \textit{Spitzer} Space Telescope observed several star-forming regions with its instruments IRAC (3.6, 4.5, 5.8, and 8 $\mu$m) and MIPS (24, 70 and 160\,$\mu$m, the last two not used in this study). IRAC1, IRAC2 and MIPS1 bands are therefore comparable to W1, W2 and W4, respectively. We computed the ratio of measured fluxes (as $\rm {WISE_{band}/IRAC_{band}}$) as a function of $\rm IRAC_{band}$ flux, where $band$ corresponds to each of the comparisons mentioned earlier. The results are shown in Fig.~\ref{fig:spitzer_wise_problem}: W1 and W2 are compatible with the corresponding IRAC photometry, whereas W4 deviates significantly from MIPS1 measurements. We found this effect to be more important for weaker fluxes, which could indicate two different issues: (1) a problem with WISE flux estimates for unresolved sources, specially important for confused regions, or (2) an important influence of extended background emission. Figure~\ref{fig:spitzer_wise_problem} and \ref{fig:ic348} illustrate these two problems in the case of the IC~348 young stellar cluster. IC~348 is a rather dense young cluster with a high and variable background level. WISE fluxes in the W3 and W4 bands were found to be incompatible with the \textit{Spitzer} ones. The full width at half-maximum (FWHM) of the W4 PSF is 11\arcsec\, (WISE All Sky Release Explanatory Supplement), considerably larger than the 6\arcsec\, FWHM of the MIPS1 PSF (MIPS Instrument Handbooks), resulting in a higher level of confusion and a greater sensitivity to variable extended background. Figure~\ref{fig:sed_wise_spitzer} shows an example of a problematic source within IC~348. If we trust the WISE photometry blindly, a mid-IR excess is found at 12 and 22~$\mu$m. The corresponding Spitzer photometry demonstrates that no such excess exists, and that the source most likely does not harbor a disk.

\end{appendix}

\end{document}